# The Cell Physiome: What do we need in a computational physiology framework for predicting single cell biology?

Vijay Rajagopal[1,*], Senthil Arumugam[2,3,4], Peter Hunter[5], Afshin Khadangi[1], Joshua Chung[1], Michael Pan[6]
Department of Biomedical Engineering, The University of Melbourne

[1] Dept. of Biomedical Engineering, University of Melbourne, Melbourne, VIC 3010, Australia
[2] Cellular Physiology Lab, Monash Biomedicine Discovery Institute, Faculty of Medicine, Nursing and Health Sciences, Monash University, Clayton/Melbourne, VIC, Australia.
[3] European Molecular Biological Laboratory Australia (EMBL Australia), Monash University, Clayton/Melbourne, VIC, Australia.
[4] Australian Research Council Centre of Excellence in Advanced Molecular Imaging, Monash University, Clayton/Melbourne, VIC, Australia.
[5] Auckland Bioengineering Institute, University of Auckland, Auckland, New Zealand
[6] School of Mathematics and Statistics, University of Melbourne, Melbourne, VIC 3010, Australia

[*] Correspondence: vijay.rajagopal@unimelb.edu.au

## ABSTRACT

Modern biology and biomedicine are undergoing a big-data explosion needing advanced computational algorithms to extract mechanistic insights on the physiological state of living cells. We present the motivation for the Cell Physiome: a framework and approach for creating, sharing, and using biophysics-based computational models of single cell physiology. Using examples in calcium signaling, bioenergetics, and endosomal trafficking, we highlight the need for spatially detailed, biophysics-based computational models to uncover new mechanisms underlying cell biology. We review progress and challenges to date towards creating cell physiome models. We then introduce bond graphs as an efficient way to create cell physiome models that integrate chemical, mechanical, electromagnetic, and thermal processes while maintaining mass and energy balance. Bond graphs enhance modularization and re-usability of computational models of cells at scale. We conclude with a look forward into steps that will help fully realize this exciting new field of mechanistic biomedical data science.

## 1.0 INTRODUCTION

Cells are the fundamental units of life. They represent the smallest unit of biology where non-living molecules transcend into living things. In multicellular organisms, single cells represent the transition point between molecular and macroscopic scales that are highly coordinated. A deep and fundamental understanding of a cell's physiological state and its adaptability is critical to better health outcomes.

We are now seeing an exponential growth in data on the cell's physiological state. New technologies report molecular state at the resolution of organelles and molecules. They can also capture dynamics at high spatio-temporal resolution across spatial scales. Building on these advances, the Human Cell Atlas[1] project aims to classify all the different cell types in the human body using single-cell molecular profiling techniques. Electron microscopy and light microscopy have also seen significant advances in the last decade. We can now visualize biological architecture and dynamics from the tissue-scale down to the single molecule inside a single cell[2-4].

However, these big datasets raise several questions: How are the different read-outs of physiological states linked? How can we extract a mechanistic understanding of a disease faster from these datasets? How can we leverage these mechanistic underpinnings for preventative or early management of disease? The answers lie in characterizing how cells integrate multi-scale chemical and mechanical signals and drive gene expression and physiological change. We need efficient means to link together these different big datasets before answering questions about how different processes integrate to define the cellular state.

Computational tools from systems biology and bioinformatics have been extensively applied to characterize cell signalling networks to date but they have their limitations. Bioinformatics [5] approaches extract correlations and patterns in protein, metabolite, and transcript content of tissues. Mechanistic interpretations from these correlations are left to tools that perform statistical comparisons of the data against curated databases. Systems biology [6] has largely applied network interaction approaches and quantitative biochemistry modelling (mass action kinetics, or Michaelis-Menten approximation). They have only recently begun to include processes that feedback to the interaction network in a dynamical manner, e.g., the role of mechanical forces through the field of mechanobiology. Additional challenges include model complexity and failure to obey physical laws. Furthermore, the fact that these signalling cues occur in spatially localized intracellular microdomains begs computational tools that also incorporate information about the 3D cell architecture.

The Physiome Project[7] is an international effort to develop mathematical and computational modelling framework to represent, understand, and predict human physiology. The framework's key principles include: (i) biophysics-based, multi-physics (e.g. coupling fluid mechanics and solid mechanics to describe coupling of blood flow and cardiac mechanics), and multi-scale modelling of physiological processes that are grounded in conservation laws of physics; (ii) models that are anatomically detailed to account for key structure-function relationships; (iii) models defined using model sharing standards; and (iv) model predictions that can be validated with experimental data, although in many cases modelling drives new measurements. The Physiome project has led to the development of detailed computational models of many organ systems including cardiovascular[8], musculoskeletal[9], respiratory[10], digestive[11], and the reproductive system[12]. These models are now routinely used to fuse different clinical imaging and omics data to characterize the patient-specific organ-level physiological state. The patient-specific models can then be used to make treatment decisions to improve patient health.

The big data explosion in single cell biology presents an exciting opportunity to adopt and extend the Physiome framework to create biophysics-based and structurally and physiologically realistic computational models of the cell - the 'Cell Physiome'. Such models would mechanistically link datasets of 3D cell architecture, molecular biology, and physiology. It would provide a framework with which to characterise the time and spatial-scales of key mechanical and chemical cues that underpin disease progression at single-cell resolution. Through Physiome standards, these models could also interface with organ-level models to gain insights on how cellular level phenomena drive organ physiology.

With this motivation for a 'Cell Physiome modelling' framework in mind, we have structured the review as follows. We first provide an overview of the types of data that are now available to measure the physiological state of single cells. We then provide three case studies covering gene transcription, mitochondrial bioenergetics, and endosomal trafficking to highlight the need for cell physiome modelling. We then review the current state-of-the-art techniques, computational models, and frameworks for modelling cell physiology. We present an introduction to bond graph modelling to address model complexity and conservation of biophysical laws **(Fig. 1)**. We conclude with a perspective on where the cell physiome field is headed and what methods and tools are needed to lead in the right direction.

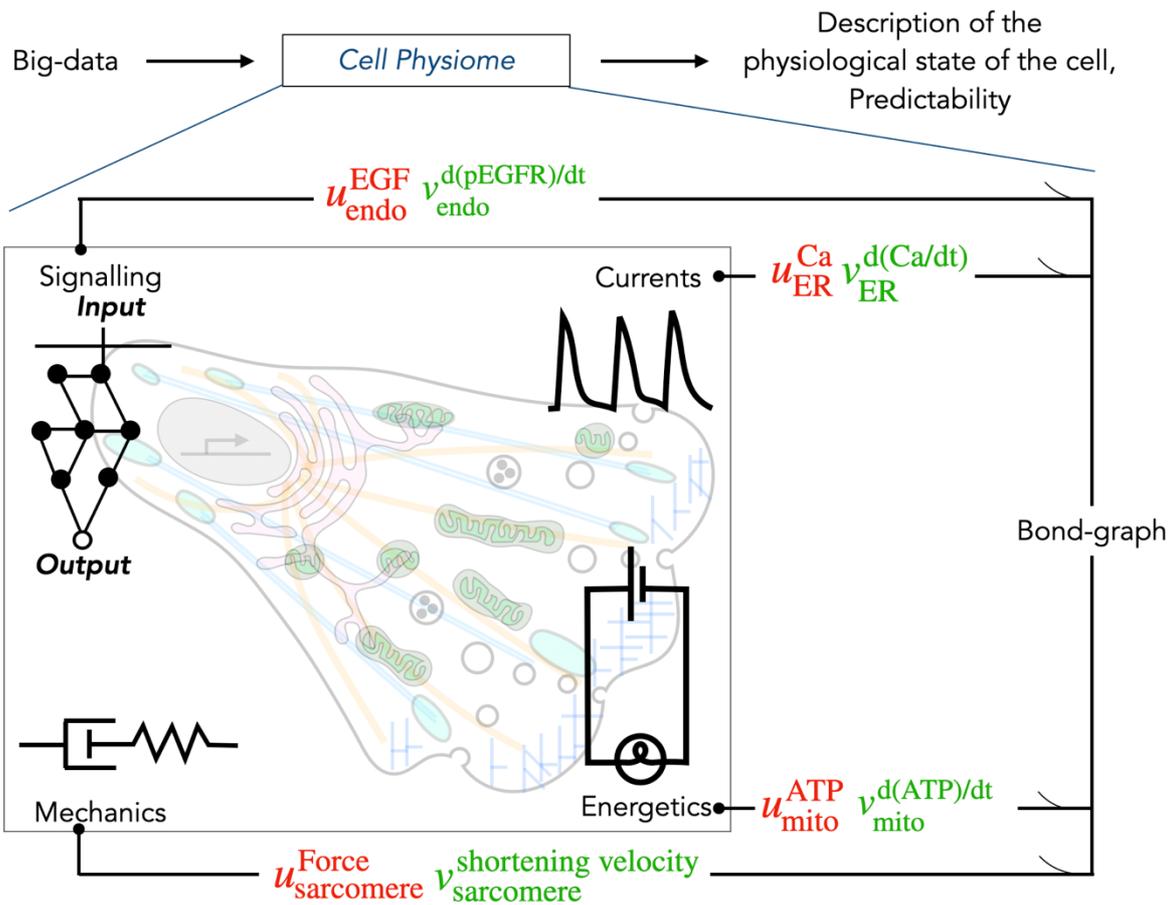

*Fig. 1:* *The Cell Physiome approach to biological data analysis: We propose to describe and predict the cellular physiological state using computational models grounded in the laws of physics. Such models would mechanistically link big datasets of 3D cell architecture, molecular biology, and physiology. Bond graph-based modelling of the potential, u, and flow, v, in different biophysical processes will efficiently integrate different sub-cellular processes into a unified model of the living cell.*

## 2.0 BIG DATA AND WHAT DOES IT ALL MEAN?

Investigating cell physiology requires careful sample preparation and precise instrumentation for measuring sub-cellular and tissue properties and processes. For example, a typical electron microscopy image requires several days of tissue fixation, resin embedding, and sample orientation before acquisition. Traditional western blotting only identifies proteins of interest, thus removing the possibility of discovering changes in expression of other proteins that may be more important or yet to be discovered. The last 20 years have seen significant advances in automation, precision, and high-throughput characterization of cellular biology. Here we highlight major advances in biological measurement techniques **(Table 1)** that generate big data.

## 2.1 Big data in number of components

Western blotting has given way to multi-omics[13] and spatial omics[5, 14, 15] technologies. Bulk omics reveal relative changes in abundance of hundreds and thousands of proteins, transcripts, and metabolites within excised tissue samples. Spatial omics techniques retain the location on the tissue from which the tissue was sampled. This enables spatial mapping of relative abundance of metabolites and proteins within cells and tissues. This spatial context is essential to infer localized changes in the tumor microenvironment[16] and other disease contexts better understand the mechanisms and the spatial signature of disease. The Human Cell Atlas Project[1] harnesses these frontier technologies to acquire a cell-level map of the physiological state of tissues in health and disease. Could we build on these efforts to bring to bear the physical laws that govern interactions and dynamics in the crowded cellular world and the constraints that these laws bring to how cells work and die?

## 2.2 Big data in time

The last two decades have also seen major technological advances in microscopy methods. The first decade of the 2000s saw the advent of confocal microscopy. The second decade saw the rebirth of light sheet-based imaging approaches. The light-sheet based modality, while negating phototoxicity, captures fast volumetric dynamics along with extending the duration for which a sample can be continuously observed. This is particularly applicable to stochastic processes, where whole cell volume observations are essential to ensure that all stochastic events within the cell volume are captured; the extended duration helps capture sufficiently large numbers of events to statistically characterize the ensemble dynamics. Light-sheet microscopy uniquely balances photon budget enabling long duration high spatio-temporal resolution imaging of live cells[17].

Another aspect of mapping complex dynamics in single cell biology is exemplified by inter-organelle interactions. Spectral detectors and calibrations for individual fluorophores are used to image six organelles simultaneously and perform linear unmixing. This hyper-spectral approach results in a six-channel image that contains a wealth of information about inter-organelle interactions[18]. In this example, the authors identify that mitochondrion most prominently interacts with the ER, and the ER-Mitochondria contact site (ERMCS), as a metaorganelle, interacts most with golgi over time. This approach holds tremendous capacity to map causal interactions between multiple components that are otherwise invisible when only a few components are imaged simultaneously. Together with technologies like DNA-Nanodevices, e.g., Voltair that measures membrane potential across organelles or calcium currents, pH etc. in live samples[19], a wealth of multiparametric data can be expected to drive the discovery of new principles in fundamental biology or in disease conditions. Better understanding of causal relationships can then provide an empirical way to design intervention strategies.

## 2.3 Big data in scale

A characteristic feature of the microscopy developments in the last decade has been that of spanning predominantly broader spatial scales and to some extent, time scales. This multi-scale nature of imaging is essential to investigate integrative physiology. For example, with the combination of adaptive optics with lattice light-sheet imaging, observations can be made at the resolution of individual organelles across a large volume of a living sample or a tissue[20]. Such datasets help uncover features and correlations across scales that are otherwise not captured by local volume imaging. For example, adjacent cells were observed to undergo cascading mitosis at 14 hours post fertilization even though synchrony of cell division is thought to be lost in Zebrafish at 3 hours post fertilization[20]. Expansion microscopy, where the fluorescent entities marking the positions of proteins of interest are physically expanded in a controlled manner, is another approach that immediately allows super-resolution imaging across a large piece of tissue[21]. The power of this technique is well demonstrated by imaging of individual synapses in very large tissue blocks. Combined with lattice light-sheet microscopy, rapid imaging of entire fly brains at nanoscale resolution can be performed in under 3 days[22]. This now allows performing unprecedented experiments and asking very important questions - Are the fly brains wired exactly in the same manner in genetically identical flies? Where do differences arise from? Do the differences arise at the level of synapses or from the process of wiring of the brain itself during development? Expansion microscopy has also furthered into smaller spatial scales in combination with single-molecule localization microscopy (Ex-SMLM) providing true molecular resolution with protein-specific labelling and preserved ultrastructural details[23]. Thus, it is evident that optical microscopy can span more spatial scales than previously possible, paving way for multi-scale imaging of fluorescently labelled species from molecular to tissue level organizations.

## 2.4 Big data in characterization

Technologies for characterizing the physical properties of cells such as electrophysiology, cell stiffness [24, 25], cell shape [26], and traction forces [27] have also increased throughput over the years. Optical tweezers can now probe viscoelastic properties at sub-cellular resolution. Microfluidic traps enable high-throughput measurement of overall cellular deformability[28] and 20 other physical parameters such as size and deformation kinetics that help to further characterize cellular state. Techniques harnessing acoustic wave patterns are now being developed to pattern cells into complex arrangements inspired by biology[29]. This same physics of acoustics can also be used to probe the mechanical properties[30].

Mechanical forces form another important aspect of characterization in single cells to developing tissues. A variety of force and tension sensors with distinct working principles are now available to use in a pertinent manner. Some examples include DNA-based force sensors that can be used to characterize focal adhesion complexes [31-33], and circularly permutated sequences of Venus and Cerulean as FRET pairs (cpstFRET) for real time measurement of mechanical stress in live cells [34, 35]. Another approach, where sensors are rationally designed using models of mechanical behaviour of synthetic unstructured polypeptides allow calibrated measurements of mechanical environment in living cells and tissues[36]. For lipid membranes, lifetime-based probes enable reporting on the membrane tension [37, 38]. Optogenetics based mechanical perturbations, which allow activation of specific mechanochemical pathways or recruitment of a mechanical regulator specifically upon light exposure, enable systematic probing of the role of mechanics in tissue development. OptoGEF and optoGAP that can activate or deactive Rho1 signalling, respectively have been used to study tissue elongation in *Drosophila*[39, 40]. Mechanistic dissections of biological processes are now possible owing to combination of optogenetics with advanced imaging approaches [41, 42] and light-activatable engineered protein modules such as PA Rac1[43]. It is expected that these approaches will gain prominence in tissue level studies.

Another aspect of big data is the characterization of underlying mechanisms that are otherwise unclear from solely looking at interactome or are ambiguous owing to inherent complexity. A medley of mechanisms may feed into a larger, macroscopic process in single cells. For example, biochemical interactions between lipids and proteins drive endosomal conversion that may in turn govern cell level transport and rates of cargo delivery to specific target organelles[44]. *In vitro* reconstitution approaches, where only the essential minimum number of components are used to recreate the phenomenology add significant value[45]. An example is the lipid kinase-phosphatase competitive reaction scheme for phosphoinositide conversion where a geometry-sensing mechanism has been suggested[46].

High-throughput electrophysiology recordings are now possible using highly sensitive microelectrodes[47] and opto-genetics techniques[48]. Molecular probes utilizing DNA hybridization are now available to measure membrane potential[19] of subcellular organelles. These precision tools now surpass voltage-sensitive dye fluorescence[49] methods that are limited in spatial resolution; fluorescence methods have so far not been capable of measuring membrane potential in organelles beyond mitochondria.

Dye fluorescence has been successfully deployed in measuring many other properties within cells including a myriad of chemical concentrations, pH, spatial distribution of proteins, temperature, and more. Combined with high-resolution microscopy techniques we now have the capability to measure the cellular physiological state with high precision.

## 2.5 Cell physiome modelling to synthesize big data understanding

With all these techniques and cellular measurements at our fingertips we are still left with the puzzle of how these measurements govern the cell's physiological state. This requires simultaneous multi-parametric measurements, which has been addressed to a very limited degree in some research and commercially available systems. But there are limitations to the number of physical probes one can insert into the cell at any given time. Some measurements, like multi-omics or electron microscopy, do not preserve the tissue sample for further measurement. These techniques also limit measurement to snapshots in time. Further, we are yet to map the biophysical changes - be it mechanical forces, chemical potential gradients, or electrical currents - that act in concert to regulate many sub-cellular processes. How do biophysical mechanisms interact spatially?

Biophysics-based computational models are powerful tools for overcoming challenges in data integration and simultaneous visualization of multiple parameters. Multi-scale computational models of organ physiology have a successful history of linking different clinical datasets and simultaneously providing biophysical insight[8]. For example, computational models of the heart integrate clinical patient data such as MRI and CT while simultaneously providing physics-based predictions of spatially varying stress, strain, and cavity pressure. In the following section we provide three case studies of where cell physiome modelling would transform our understanding of cell physiology.

## 3.0 THE NEED FOR CELL PHYSIOME MODELLING

We present three specific examples of important cellular processes that would benefit from cell physiome modelling **(Fig. 2)**. We first introduce the importance of modelling sub-cellular organization of calcium release channels to study gene transcription in cardiac hypertrophy. We then highlight recent studies that point to a complex interplay between spatial organization of mitochondria, bioenergetics and many other processes that interact with mitochondria. We present a final case study on endosomal trafficking, a critical discrete intracellular process by which external signals modulate the cell's physiological state.

### 3.1 Cell signaling compartmentation: calcium in cardiac hypertrophy

A healthy heart is essential to providing and maintaining the cardiac output required to meet short- and long-term hemodynamic demands of the body. When subject to chronic cardiac stress while carrying out this function, the heart responds by undergoing cardiac hypertrophy, which manifests largely through the growth of cardiomyocytes. This growth compensates for the increased load brought about by the cardiac stress, thereby adapting the heart to its long-term exposure. [50]. Calcium ($Ca^{2+}$) plays a key signalling role in this matter **(Fig. 2a)**.

Calcium ($Ca^{2+}$) is a critical node in signalling pathways responsible for cardiomyocyte growth that leads to cardiac hypertrophy. Specifically, $Ca^{2+}$ is implicated in signalling pathways that target pro-hypertrophic transcription factors nuclear factor of activated T-cells (NFAT) and myocyte enhancer factor 2 (MEF2)[51]. The enhanced activation of which is associated with cardiac hypertrophy induced by pathological conditions[52, 53]. At the same time, $Ca^{2+}$ is also central to the excitation-contraction coupling (ECC) process that underlies the heart's pumping[54]. Central to this process is the flood of $Ca^{2+}$ released from the sarcoplasmic reticulum (SR), the cardiomyocyte's intracellular $Ca^{2+}$ store, via ryanodine receptors (RyRs) into the cytosol and its subsequent reuptake back into the SR. The resultant rise and fall of cytosolic $Ca^{2+}$ concentration (also called cytosolic $Ca^{2+}$ transient) facilitates the contraction and relaxation of the cardiomyocyte.

How $Ca^{2+}$ signals activate pro-hypertrophic transcription pathways amid ECC-associated $Ca^{2+}$ transients inundating the cytoplasm of the cardiomyocyte has been a longstanding question [55]. In this regard, experimental evidence reporting the necessity and sufficiency of nuclear Ca$^{2+}$ transients in activating NFAT and MEF2 [56, 57] engendered a hypothesis that the spatial compartmentation of Ca$^{2+}$ signals (to the nucleus and to the cytosol) may, to some extent, explain how Ca$^{2+}$ can simultaneously regulate transcription and contraction. Notably, these nuclear Ca$^{2+}$ transients are potently regulated by inositol 1,4,5-trisphosphate (IP$_3$) receptors (IP$_3$Rs)[56, 57], another family of Ca$^{2+}$ channels that are activated downstream of hypertrophic stimuli that engage G-protein coupled receptors (GPCRs).

The cytosolic and nuclear compartmentation of Ca$^{2+}$ signals may be permitted through the spatially heterogeneous expression of the two Ca$^{2+}$ channels (RyRs and IP$_3$Rs), as well as their differential regulation within the cardiomyocyte. RyRs are exclusively expressed on the SR outside the nuclear region; IP$_3$Rs are predominantly expressed in the perinuclear region and, to a smaller extent, colocalize with RyRs on the SR[58]. RyRs and IP$_3$Rs are also activated by different ligands. While cytosolic Ca$^{2+}$ can act as the sole activating ligand for RyRs[59], IP$_3$Rs require IP$_3$ in addition to Ca$^{2+}$ for its activation[60]. Avenues and mechanisms for crosstalk between cytosolic and nuclear Ca$^{2+}$ signals are also less explored. While it is reported that ECC-associated cytosolic Ca$^{2+}$ transients are insufficient to activate NFAT and MEF2[56, 57], there is compelling evidence to show that nuclear Ca$^{2+}$ can be regulated by cytosolic Ca$^{2+}$ through passive diffusion and local adjacent signalling events[61]. Nuclear Ca$^{2+}$ transients were also observed to occasionally trigger cytosolic Ca$^{2+}$ waves[62]. Moreover, enhanced cytoplasmic IP$_3$R activity was also shown to modulate ECC-associated Ca$^{2+}$ transients and increase the frequency of spontaneous Ca$^{2+}$ release events[58]. Together, these observations not only highlight the possible influence nuclear and cytosolic Ca$^{2+}$ signals have over each other despite their apparent compartmentation but also the modification of those signals due to different regulation mechanisms of RyRs and IP$_3$Rs.

The overall effect of this inter-channel to inter-signal Ca$^{2+}$ crosstalk may have significant implications on the progression of the cardiomyocyte's hypertrophic response. This can be best investigated at a whole-cell level with concurrent ECC and hypertrophic signalling. However, the initiation and termination mechanisms of nuclear Ca$^{2+}$ transients are less understood. Like ECC-associated cytosolic Ca$^{2+}$ transients, a fundamental understanding of IP$_3$R-mediated Ca$^{2+}$ release in the perinuclear region may be required to elucidate the elementary events constituting nuclear Ca$^{2+}$ transients and the effects external perturbations may have on it. Accounting for the role of subcellular structure in this aspect will be imperative. The topology of the nuclear envelope, its proximity to organelles in the perinuclear region, and the corresponding arrangement of Ca$^{2+}$-handling proteins on its membrane are all factors that may impact local Ca$^{2+}$ diffusion, nuclear $Ca^{2+}$ transient formation, and its apparent segregation from cytosolic $Ca^{2+}$ signals. Furthermore, the respective contribution of $IP_3R$ and RyR $Ca^{2+}$ release to ECC and hypertrophic signalling, which forms the basis for cytosolic and nuclear $Ca^{2+}$ crosstalk, are experimentally indistinguishable. Computational modelling is an alternative to addressing these limitations through higher resolution measurement and tracking of intracellular spatiotemporal $Ca^{2+}$ dynamics. Specific to $Ca^{2+}$ signalling in cardiac hypertrophy, a cell physiome approach to modelling the cardiomyocyte would account for the intracellular chemical and structural interactions of $Ca^{2+}$ while ensuring continuity between protein-level and cellular-level simulations. With this, how subcellular structure and RyR-$IP_3R$ interaction and placement lead to spatially segregated yet interacting $Ca^{2+}$ signals to simultaneously regulate ECC and cardiomyocyte growth can be fully elucidated.

## 3.2 Spatiotemporal energetics: mitochondria organization, dynamics, bioenergetics and more.

Adenosine triphosphate, or ATP, is the energy currency inside living cells. A mitochondrion is an efficient source of ATP as it houses the machinery for oxidative phosphorylation. Mitochondrial content correlates highly with the energy demands of a cell, making up ~50% of the volume of cardiomyocytes[63], the cell-type with the highest energy demands in the human body. Within the last decade our understanding of mitochondria as static organelles producing ATP on demand have changed. Understanding the interconnection between mitochondria shape, organization, and dynamics and their impact on a variety of cell processes present an exciting frontier for molecular biology and cell physiome modelling research.

Mitochondria shapes, size, and organization correlate with energetic and physiological state[64, 65]. Cardiomyocytes exhibit small, fragmented mitochondria populations in many disease conditions like diabetic cardiomyopathy and ischemic reperfusion injury[66]. Mitochondria exhibit distinct differences in shape and size within a neuron's soma, dendrites, and axon, owing to different energetic demands within these compartments of the cell. 3D electron microscopy has revealed the intricate networks that mitochondria form within the cell[67]. These networks may also act like a power network; recent experimental data suggest that the mitochondrial membrane potential is transmitted like an electrical current across a mitochondria powerline[68]. This brings to question our prior approach to studying mitochondria within a cell as two populations[69]: those directly underneath the cell membrane and those deep within the cytoplasm.

Questions remain about what the network topology reflects about the energetic capacity of the cell.

The dynamic processes of other organelles such as ER and lysosomes play an important role in the cell's regulation of its mitochondria[70]. The maintenance and modulation of mitochondria are 'encoded' in the energy consuming processes of organellar motility, membrane sculpting enzymes, and cytoskeletal dynamics themselves. ER forms contact sites with mitochondria and participates in the process of mitochondrial fission[71] as well as exchange $Ca^{2+}$[72], while lysosome-mitochondria interactions are critical for homeostasis and metabolism[73]. Lysosome-mitochondria interactions are also important for calcium dynamics in the mitochondria that are signals for cellular stress[74]. The inter-organelle interactions underscore the need for integrating the spatial arrangement of multiple organelle types and the multiple processes residing in them. Such structure-function sub-cellular models will help elucidate how the cell utilizes its components for more than one purpose. Moreover, functions and consequences of calcium currents that occur between the organelles are yet to be fully understood. The positioning of organelles and location dependent dynamics and regulation is an overlooked issue that is expected to gain prominence with higher-resolution data (in space, time, and number of components).

Mitochondrial dynamics and their location are revealed to be significant for cellular processes or reflect the state of the cell. For example, mitochondrial locations are implicated in cell migration or epithelial to mesenchymal transition[75]. Small molecule inhibitors of PI3K, a cancer-relevant target, reprograms the cell, and interestingly involves trafficking of mitochondria to cell periphery, supporting tumor cell invasion[76]. As another example, mitochondrial locations and local metabolism are emerging to be significant for synaptic plasticity[77]. Mitochondria are shown to be tethered by cytoskeletal elements within stable spatial compartments and provide energy for local translation within the spatially confined boundaries[78]. Synapse formation is a complex multi-scale process, and cancer relevant cell reprogramming that influences mitochondrial localization represent complex interlinked processes that take place in living cells and emphasizes the need for a quantitative integrative approach that enables predictability.

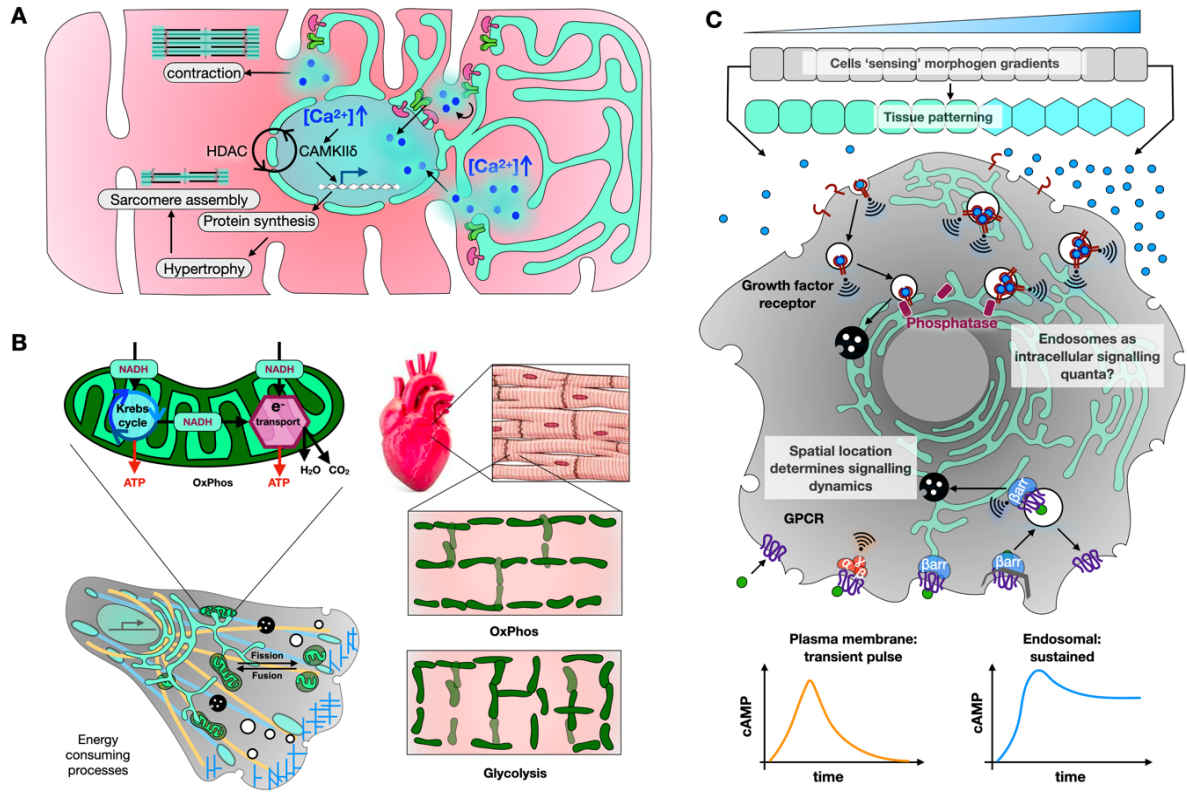

*Fig. 2:* *(A) Cardiac hypertrophy is regulated by calcium release events at discrete nanodomains within the cardiac dyad and the nuclear envelope. These release events are regulated by the opening and closing of ryanodine receptors and IP3 receptors, which have different sensitivities to calcium. (B) Mitochondrial OxPhos drives energy consuming processes in living cells. Cells dependent on OxPhos have grid-like mitochondrial organization. Glycolytic cells have perpendicularly oriented (to the myofibrils) mitochondrial networks. (C) Endosomes as units of signalling. (Top) Interpretation of gradients by single cells leading to different outcomes. (Bottom) Transient and sustained signalling upon agonist binding to GPCRs dependent on location of active GPCR receptors.*

### 3.3 Cell signal processing: endosome trafficking

Cells interpret signals via receptor proteins on the cell surface and send it forwards via various signalling pathways. Various detection schemes are required to decode a plethora of time-varying signals or ligand multiplicity for the same receptors. In this sense, a single cell acts like a computer program, where there is an input (for example, ligands binding to receptors), processing (yet unknown), and an output (change in the metabolic state of the cell or cell fate decisions etc.). How a cell processes the inputs and converts them into outputs remains an important unsolved question[79, 80]. For example, how genetically identical cells interpret gradients in growth factors to decipher positional information to determine cell fates or tissue patterning is not yet known. Furthermore, multiple signals are received by the cells that define cell fate. How are cell fate decisions made by cells that experience chemically similar, but distinct concentrations of multiple signals?

In the case of some growth receptors, distinct growth factors can bind to the same receptor, and yet evince different cell fate outcomes such as proliferation or migration[81]. It is now widely accepted that endocytosis of most receptors does not result in deactivation, and that deactivation could be occurring at specific inter-organelle interaction sites[80]. Whilst some key aspects and molecular players regulating endocytosis and intracellular trafficking via endosomes are known, the mechanism a cell uses to interpret these signals remains poorly understood **(Fig. 2c)**. Compared to ~1500 µm$^2$, endosomes constitute about 20000 µm$^2$ with more precisely defined discreteness in compartmentalization that at the same time are interchangeable and highly regulated.

An exciting idea put forward by Zerial and colleagues was that the number of phosphorylated EGFRs per endosome are regulated and not the total number of EGFRs per endosomes. They hypothesize that the regulation of phosphorylated EGFR clusters could be derived by feedback loops between enzymes that regulate phosphorylation-dephosphorylation cycles that are yet to be demonstrated[82]. Stanoev *et al.* propose a principle of temporal information processing based on dynamic transient memory. They propose that the recycling pathway provides a mechanistic means to maintain critical receptor concentration at the plasma membrane through a fluctuation sensing system for a network system organized at criticality[83]. Furthermore, recent evidence suggests that signalling, calcium currents, and modulation of localization of endosomes to the perinuclear region can all be integrated in a concerted fashion. Taken together, these studies strongly suggest that the endosomal network is the logistical means of the cell to 'process' the input signals via modulation of receptor distribution between the cellular exterior and specific intracellular reaction niches (e.g., ER-lysosome contact sites[84]) as well as the membrane of the endosomes provide a rich substrate for complex non-linear interactions, e.g., reaction-diffusion to take place[79].

Directional endosomal transport is managed by the two families of motor proteins that execute transport along microtubules. The cross-organellar interactions that are central to modulate receptor activity result from the multiple collisions created by the transport, and the active nature of the cytoplasmic constituents[85]. How are desirable interactions sustained over non-desired ones? What manages the timeliness of these seemingly noisy processes are only beginning to be unravelled. It is plausible that emergent mechanisms arise out of local molecular interactions and/or motility driven cross-organelle interactions and overcome the stochastic nature of the contributing processes that build up macroscopic deterministic behaviours.

An exciting set of measurements, enabled by DNA-based nanodevice that measures membrane potential of endocytic vesicles and the trans-Golgi network in live cells revealed that membrane potentials of early endosomes, late endosomes, and lysosomes to be +153 mV, +46 mV and +114 mV, respectively[19]. Interestingly, trans-Golgi network was measured to have a membrane potential of 121 mV, while recycling endosomes displayed +65 mV, like the plasma membrane, suggesting similar electrical characteristics[19]. Together with inter-organelle contact sites, which have been shown to be 'electrically coupled' for transient temporal bursts, exemplified by calcium currents[74], it remains to be uncovered what are the consequences of membrane potentials and the bursts of currents through them. Whilst not clear at the resolution of single endosomal entities, receptor activation that leads to calcium transients have been shown to perturb the pre-existing steady state of an endosomal adaptor, APPL1, to result in increased cytosolic contractions enabling preferential binding to activated EGFRs resulting in cohort trafficking[86]. Whether transient local currents at individual organelle level re-configure the protein biochemistry remains to be elucidated.

Light-sheet-based high-resolution techniques are increasingly improving on the resolution and are expected to cater to the specific requirement of volumetric imaging with high spatio-temporal resolution with maximum photogentleness to capture stochastic events that build up the macroscopic process[17]. Multiple-species and their dynamic characteristics are now possible to accurately measure across the scale of individual endosomes to whole-cell or tissue level[20, 87]. High-resolution multiscale imaging will enable more complex data to be acquired that can feed into a framework that simultaneously integrates stochastic processes in endosomal trafficking and signalling with quantitative description of receptor activity and its downstream effectors. The precise roles of endosomes as signalling quanta and the mechanisms of cellular interpretation of bulk ligand characteristics remains to be elucidated. In summary, endosomal trafficking is a critical and complex process in cellular physiology that has not undergone a quantitative and integrative modelling treatment. New discoveries are being made directly implying endosomal trafficking regulators in cancer [88]. Cell Physiome based approaches will uncover mechanistic aspects of endosomal regulators in cancer.

## 4.0 STATE-OF-THE-ART TOOLS FOR CELL PHYSIOME MODELLING

The examples outlined in Section 3 are only a small sample of biological processes that would benefit from cell physiome modelling. Here we review the state of the art in integrated modelling of cellular processes and their 3D environment. We highlight critical gaps in tools for advancing cell physiome modelling. We introduce a new approach to creating biophysics based computational models of the cell physiome using bond graphs.

4.1 Mixed-compartment models of biochemical reaction networks

Although cell physiome modelling is an emerging area of biomedical data science, it has historical precedent in the field of systems biology. Systems biology started from simple, lumped parameter models aiming to explain specific aspects of cell physiology, for example the Michaelis-Menten model of enzyme kinetics [89] or the Hodgkin-Huxley model of electrophysiology [90]. The recent boom in biological data has made it possible to develop more complex models of cellular biochemistry. In the context of cardiac cell biology, mathematical models are now able to integrate the processes of electrophysiology, signaling, bioenergetics, and contraction[91, 92]. These models can be used to understand the mechanisms of heart disease and to assess drug toxicity[93]. There have also been efforts to generate whole-cell models that account for every biomolecule in the cell. Such models have great potential in directing personalized medicine in clinical settings and informing design in synthetic biology. The first whole-cell model was developed for the *Mycoplasma genitalium*, a relatively simple bacterial organism [94]. Ongoing efforts are being put into developing whole-cell models of more complex organisms such as *Escherichia coli* and humans [95, 96]. However, despite the amount of biological detail that these models incorporate, complex models of biochemistry rarely incorporate space as an explicit variable and are therefore unable to model the inhomogeneous spatial distribution of chemical species throughout the cell. This motivates the need for 3D models that can account for both the complexity of cellular biochemistry and spatial heterogeneity.

4.2 3D models of cell physiology

3D cell modelling has been harnessed in several biomedical research fields already, albeit modelling only a limited number of processes. We highlight some state-of-the-art 3D cell models as a window into the future of cell physiome modelling. These models also point to bottlenecks in modelling techniques that present exciting opportunities for innovation that will transform biomedical data science.

The deformability of the red blood cell is critical to its function of oxygen transport through the vascular system. Finite element modelling [2, 26] and coarse-grained modelling[3] techniques enable biophysically detailed computational modelling of red blood cell mechanics. These models have been used to study the influence of red blood cell stiffness[26, 97] to blood transport [98] in health and disease. These computational models are particularly beneficial in dissecting confounding factors, like cell stiffness and cell shape, that govern red blood cell transport[26].

Cellular forces and mechanobiology[99] play an integral part in cellular physiology. Cell motility is one such physiological process that governs tissue development and cancer metastasis. Cell motility involves complex interplay between chemical signals and mechanical forces at the leading and trailing ends of the cell[100]. Migrating cells undergo a variety of shape changes during motility, with mesenchymal (rod-like), and amoeboid shapes commonly adopted [101]. A variety of computational approaches like coarse-grained molecular dynamics methods[102], level set methods [103], continuum[104, 105] and discrete, mass-spring type[106] methods have been used to gain insights on the interaction between mechanical forces and chemical signals at the integrin-adhesion-cell cortex interface during migration.

Integrated modelling has a long history in neuroscience and muscle physiology. Starting with the Hodgkin-Huxley [90] model of action potential propagation in the squid giant axon, many non-spatial computational models have been developed of the biophysics of intracellular calcium signalling [107], action potential propagation [108], mitochondrial energetics[109], and sarcomere cross-bridge kinetics[110]. More recently these models have been mapped onto spatially detailed 3D computer models of the sub-cellular anatomy to uncover the role of sub-cellular organization on cell physiology. Today, we can generate 3D geometric models of cell architecture that are derived directly from 3D electron microscopy and light microscopy data [111-114]. Organization of mitochondria, acto-myosin sarcomeres and ion channel distributions can be explicitly represented in these models. Computational meshes derived from these models can then be used to simulate calcium signalling, spatiotemporal bioenergetics, and mechanics. For example, using a computational model derived from 3D electron microscopy data, we quantified the independent effect that the spatial organization of mitochondria have on sub-cellular metabolite distributions, cross-bridge kinetics, and force production[63]. This approach has recently been applied to a whole cell kinetic model of a genetic minimal bacterial cell (JCVI-syn3A) that was derived from cry-electron tomography data[115].

## 4.3 Deep learning models for microscopy image segmentation and 3D modelling

A major bottleneck in developing integrated models of cell physiology is the generation of detailed models of cell architecture. Serial block-face electron microscopy provides information on sub-cellular architecture at 10-100 nanometre spatial resolution. But annotation and segmentation of key structures is a tedious task. Deep learning models from computer vision present a much more efficient approach to microscopy image processing. Based on our experience, manual annotation of 700 electron tomography slices may require 6-12 months of an individual's working hours for manual annotation. With deep learning models, one needs to only annotate one or two slices with care. After training, 700 remaining images will be segmented by the deep learning model in a couple of minutes.

Many deep learning models have been developed or adopted from other applications for image segmentation within 2D or 3D setting, including VGG [116], ResNet [117], V-Net [118], U-Net [119], CDeep3M [120], and EM-Net [121] to name a few. Our benchmarking software EM-Stellar [122] showed that no one convolutional neural network performs well for all data despite having complex feature layers or increased number of parameters. The performance of any network is highly dependent on the structures being segmented and the inherent properties within the data being analyzed. Nevertheless, structures like mitochondria are segmented relatively easily. In contrast, components like the endoplasmic reticulum or Z-disks [123], which make up typically 1%-2% of the cell volume present a data imbalance problem that has so far been best resolved by labelling and segmenting all other components within the cell simultaneously. Recent segmentation[1] of a whole cell from a 3D focussed ion-beam electron microscopy dataset is a testament to the power of deep learning in segmenting electron microscopy data at high throughput.

Electron microscopy is effectively untargeted visualization of the cell's content at low contrast. Super-resolution microscopy provides targeted visualization of labelled proteins. Correlated light and electron microscopy (CLEM) is actively being pursued to visualize the spatial relationships between proteins and the surrounding architecture within the cell. However, efficient ways to extract the spatial relationships from 3D data-CLEM or otherwise-and the variability within cell populations are sorely lacking. An exciting avenue for advancing deep learning is microscopy image fusion to tackle this problem. Generative adversarial networks are a class of deep neural nets that learn a probability distribution of image intensities and other higher order features. The Allen Cell Institute trained a GAN using 4,400 fluorescence microscopy images of stem cells per labelled structure[2]. The trained model can generate new synthetic cell images that reflect the joint probability distributions of different labels along with the 3D cell architecture. As such, the generative adversarial network (GAN) fused different labels in the same modality. A similar approach could be taken to computationally fuse EM and light microscopy data as well.

But currently GANs are computationally intensive to train, making the pixel resolution of images on which they can be trained low. GANs also need large datasets for training - a perennial problem in artificial intelligence. Clever computational strategies should be investigated to extract probabilistic models of cell geometry. For example, we recently applied a 2D StyleGAN [124] to extract a pseudo-3D probability distribution of cardiomyocyte architecture from 3D electron microscopy data [125]. We harnessed the high z-resolution of the 3D electron microscopy data to train the 2D GAN on the individual image slices of the volume stack. We then generated synthetic 2D images and re-ordered them in a way that minimized the overlapping error between consecutive image slices.

4.4 Tools for model reproducibility, sharing and modularizing

A key requirement for coping with the complexity of multiscale modelling is the use of computational modules: self-contained model components, with well-defined functionality, that can be imported into higher level or more complex models [7]. The [Physiome Project](#) has developed the CellML [126] and FieldML [127] standards to help ensure error free and reproducible models, and a Physiome Model Repository ([PMR](#)) has been established to make curated and annotated models covering a very wide range of biological processes, freely available to the modelling community[128]. A freely available and open-source simulation software programme ([OpenCOR](#)) for creating, editing and running CellML models has also been developed [129].

It is noteworthy that of the current ~1000 models in PMR, coded up from peer-reviewed published models, only about 10 (1%) did not require correction or additional information from the authors in order to produce a working model. Therefore, to provide an additional incentive for modelers to develop reproducible models, a new open access journal called *[Physiome](#)* has recently been launched. A *Physiome* paper is secondary to a primary publication that deals with the experimental data and rationale for the model and is intended solely to ensure that the model is available to others in a reproducible and appropriately annotated form.

Tools for model reusability have also been developed outside of the Physiome Project. Notably, the Systems Biology Markup Language (SBML) is a standard for describing a wide range of models in biochemistry [130], with many of these models being housed in the BioModels database [131]. In addition to describing the model itself, model reproducibility requires standards for simulation protocols; this is addressed by the Simulation Experiment Description Markup Language (SED-ML) [132].

Ongoing work is aimed at addressing limitations of current standards. Firstly, we need to ensure that the link between published models and data is clear. While current standards can describe the equations of a model unambiguously, it is difficult to systematically describe how parameters are fitted to data. Secondly, we need to record model provenance, that is, document the history of how models are updated as new data and observations become available. These limitations can be addressed by taking inspiration from continuous integration in software development. When a model is initially published, a protocol for fitting the model to data should be made available. This could then be used to write up a series of unit tests to ensure the model is consistent with experimental data. Finally, when a model is either updated or used within another model, there should be a link to the original model so that the link to data is preserved. The Cardiac Electrophysiology Web Lab aims to address these issues in the context of cardiac electrophysiology [133]. For more general systems biology modelling, new concepts such as 'ModelBricks' have been suggested to ensure that the context and history of a model is not lost [134].

## 4.5 Bond graphs: ensuring energy conservation and enhancing modularity and multi-physics integration in cell physiome models

For models to be integrated in a physically consistent manner, the models themselves need to comply with the laws of physics. Unfortunately, however, many of the papers currently available on PMR in a reproducible form do not obey the laws of physics. A reproducible model, even one whose parameters have been fitted to experimental data, does not necessarily satisfy the conservation laws of physics [135]. The Physiome Project has therefore also begun the development of a mathematical framework to help ensure that reproducible models of biological processes do indeed also obey the laws of physics.

In addition to *energy* (measured in Joules) and *time* (seconds), there are only five types of quantity in all of physics and hence physiology (cell, tissue, and organ, but not sub-atomic): these are *length* in meters (or its 2D and 3D equivalents, area in m$^2$ and volume in m$^3$), *electric charge* (Coulombs or C), *luminous intensity* (candela or cd), *amount of substance* (moles or mol), and *entropy* (e). These last four are defined by counting the number of microscopic entities in that quantity: i.e., the number of electrons, photons, atoms, or probable states, respectively. The conservation laws of physics are mathematical statements about the macroscopic quantities, e.g. conservation of energy, mass, or charge.

At a physiological level all physical processes fall into just four categories, listed here with their corresponding units: (i) *mechanics* with units of metres (m) or $m^2$ or $m^3$; (ii) *electromagnetics* with units of Coulombs (C) and candela (cd); (iii) *biochemical processes* with units of moles (mol); and (iv) entropically driven *heat flow*, with units of entropy (e). Almost every physiological mechanism involves all four physical processes. For example, cells involve multiple biochemical processes, often with electrically charged species, and are mechanically active, or at least sensitive to stretch, and of course involve heat production. We will show now that compartmental (lumped parameter) models can be represented by the 'bond graph' approach, which uses conservation of power as the governing principle [136]. Formulating models in bond graph form is especially advantageous when the model needs to account for energy moving between these four types of physical processes. Finally, although we do not discuss this further, the bond graph approach we will discuss here for describing compartmental cellular mechanisms is, in fact, the 0D (zero-dimensional) version of a more general approach called port-Hamiltonians [137] that includes equations based on the spatial gradients of the continuous fields used in 3D models.

The starting point for bond graphs is to consider the quantities involved in power transmission. If a quantity $q$ has units of m, C, cd, mol or e, the rate of flow of that quantity $v$ has units $m.s^{-1}$ (or $m^2.s^{-1}$, $m^3.s^{-1}$), $C.s^{-1}$, $Cd.s^{-1}$, $mol.s^{-1}$ or $e.s^{-1}$, and the potential difference $u$ that drives that flow has units Joules per unit of $q$: i.e. $J.m^{-1}$ (= force in Newtons) or $J.m^{-3}$ (= pressure or energy density in Pascals), $J.C^{-1}$ (= electric potential in volts), $J.cd^{-1}$ (electromagnetic potential), $J.mol^{-1}$ (chemical potential), and $J.e^{-1}$.(= temperature in Kelvin). The product of the driving potential $u$ and the flow $v$ is therefore always power in $J.s^{-1}$.

Now consider the flow of a quantity $q$, represented by a 'bond', as shown by the blue line in **Fig. 3a**, that is labelled with its flow $v$ and energy gradient $u$. The product of the flow $v$ ($q$ per second) and its driving energy gradient $u$ (Joules per $q$) is always power ($J.s^{-1}$). Since the laws of physics require power to be conserved, stored or dissipated, we can define two situations when these bonds meet without loss of power ($\sum u.v = 0$): one in which $u$ is constant, shown as a 'zero node' ('0:node') in **Fig. 3b**, and one in which $v$ is constant, shown as a 'one node' ('1:node') in **Fig. 3c**. The zero node is therefore associated with mass conservation ($\sum v = 0$) and the one node with energy conservation ($\sum u = 0$). For electrical circuits, the zero and one nodes correspond to Kirchhoff's current law and voltage law, respectively. For mechanics these represent mass conservation and Newton's laws. For biochemical systems, where $q$ is the quantity of a specific chemical species, they correspond to mass conservation for that species, and specific energy conservation, respectively.

Note that we depict quantities $q$ and flows $v = \frac{dq}{dt}$ in green, and energy variables $u$ in red, in order to highlight the fundamental distinction between them: the first two denote physical quantities and their rates of change; the second denotes a potential. These are also called, kinematic variables and kinetic variables, respectively.

Along with energy flows, we also need to consider energy storage and dissipation. For biochemical systems, energy storage is achieved by dissolving a solute in a solution, and dissipation (conversion to heat or entropy) is associated with a reaction. In a mechanical system, energy can be stored statically with a stretched spring or dynamically with inertial mass, and energy is dissipated in a viscous 'damper'. In an electrical system, energy can be stored statically with a capacitor or dynamically with an inductive coil, and energy is dissipated (converted to heat) with a resistor.

Static energy storage is defined empirically as a relationship between $u$ and $q$. For example, the force, voltage or temperature needed to stretch, charge or heat a spring, capacitor or thermal mass, respectively, is $u = q/C$, where $C$ is the spring compliance, electrical capacitance, or thermal capacitance, respectively. For biochemical systems the constitutive relation for static storage is given by the Boltzmann thermodynamic relation

$$u = RT\ln(Kq) \quad \text{(J.mol}^{-1}\text{)}, \tag{1}$$

where $RT \approx 2.5$ (kJ.mol$^{-1}$) at body temperature, and $K$ (mol$^{-1}$) is a thermodynamic constitutive parameter for the chemical species [138].

Dissipation is defined empirically by specifying the relationship between the flow $v$ and the potential $u$. For most physical systems this is a relationship between the flow through the dissipator and the potential difference across it (for example, Ohms' law in electrical systems). However, for biochemical processes it is found that the Marcelin-de Donder formula generally provides a good empirical fit [138]:

$$v = \kappa\left(e^{u_1/RT} - e^{u_2/RT}\right) \quad \text{(mol.s}^{-1}\text{)} \tag{2}$$

where $\kappa$ (mol.s$^{-1}$) is the rate parameter for the chemical species flowing through this reaction at velocity $v$ (mol.s$^{-1}$). This is illustrated in **Fig. 3d.**

Note that substituting the Boltzmann relation (1) into the reaction scheme (2) yields the familiar mass action relation

$$v = \kappa\left(e^{u_1/RT} - e^{u_2/RT}\right) = \kappa(K_1 q_1 - K_2 q_2) = k^f q_1 - k^r q_2,$$

where $k^f = \kappa K_1$ and $k^r = \kappa K_2$ are the forward and reverse *kinetic constants*.

This empirical relationship between kinematic and kinetic variables is called a *constitutive relation* and is an empirically measured property of the material, not a physical law!

One final energy-conserving element is often used in bond graph models. The *Transforming Factor* (TF) is used when power is converted within or between the physical domains mentioned above. For example, the flow of electrical current, driven by a difference in electrical potential is often linked with a biochemical process, such as the diffusion of ions down a concentration gradient. Or, as we illustrate below, the mechanical force developed by myofilament cross-bridges affects the unbinding of calcium from troponin-C. Examples of transformation within physical domains include electrical transformers, mechanical levers, and chemical reaction stoichiometry.

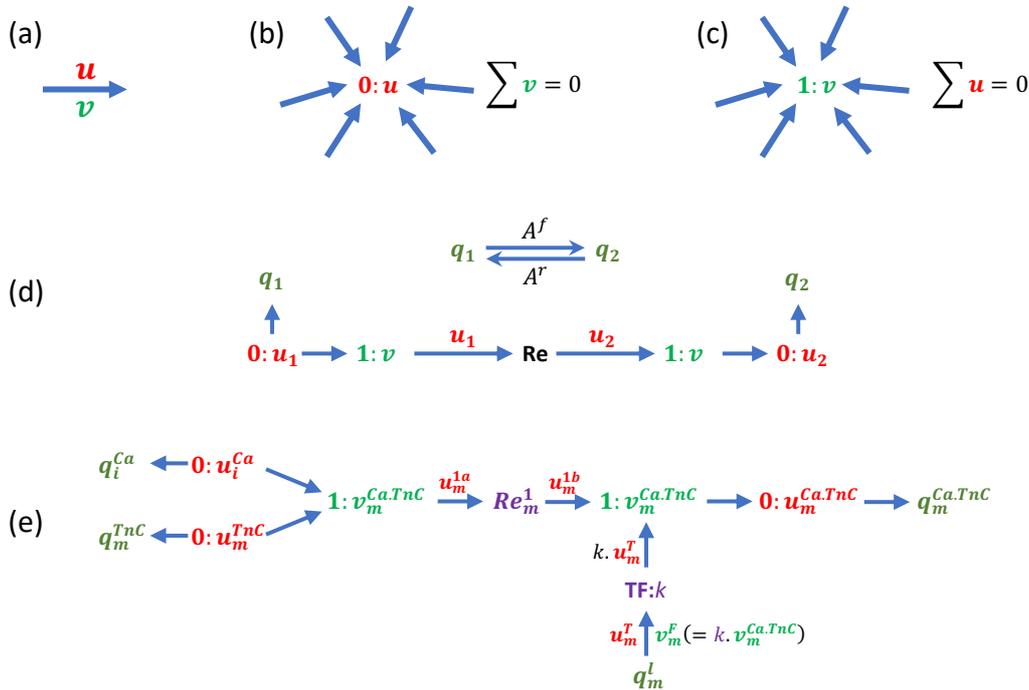

*Fig 3:* (a) A bond representing power transmission with co-power variables $u$ and $v$. (b) A 0:node junction with a common potential $u$ and the corresponding mass or charge balance $\sum v = 0$. (c) A 1:node junction with a common flow $v$ and the corresponding energy balance equation $\sum u = 0$. (d) A bond graph representation of a simple reaction. (e) A bond graph representation of the reaction between calcium (Ca) and troponin-C (TnC).

As an illustrative example of the application of bond graphs, consider the binding and release of calcium with Troponin-C, where the rate of unbinding depends on the mechanical force generated in the myofilaments. The release of Ca from TnC is slowed when myofilament tension is high, and this is represented by the power flow shown at the right of the reaction in **Fig. 3e**. The power associated with the product of force $u_m^F$ and myofilament velocity $v_m^F$ below the TF term is equal to the power associated with the product of chemical potential $ku_m^F$ and molar flow $v_m^{Ca.TnC}$ shown above the TF term. Note that the binding of Ca to TnC is much faster than the release step and does not depend on myofilament force. The equations associated with the bond graph model are generated by considering mass balance at the 0:nodes (shown in red, along with the chemical potentials $u_{location}^{species}$ in **Fig. 3e** and energy balance at the 1:nodes (shown in green, along with the flows $v_{path}^{species}$). Every 0:node has a corresponding storage term $q_{location}^{species}$ for that molar species.

Mass balance at the 0:nodes gives:

$$0: u_i^{Ca} \qquad \frac{d}{dt}q_i^{Ca} = -v_m^{Ca.TnC}$$

$$0: u_m^{TnC} \qquad \frac{d}{dt}q_m^{TnC} = -v_m^{Ca.TnC} \qquad (3)$$

$$0: u_m^{Ca.TnC} \qquad \frac{d}{dt}q_m^{Ca.TnC} = v_m^{Ca.TnC}$$

where $q_i^{Ca}, q_m^{TnC}, q_m^{Ca.TnC}$ (mol) are the chemical species involved in the reaction, and $v_m^{Ca.TnC}$ (mol.s$^{-1}$) is the flow of these quantities through the reaction.

Energy balance at the 1:nodes gives

$$1: v_m^{Ca.TnC} \qquad u_m^{1a} = u_i^{Ca} + u_m^{Tn}$$
$$1: v_m^{Ca.TnC} \qquad u_m^{1b} = u_m^{Ca.TnC} - k.u_m^F \qquad (4)$$

where $u_m^{1a}, u_i^{Ca}, u_m^{Tn}$ (J.mol$^{-1}$) are chemical potentials and $u_m^F$ (J.m$^{-1}$) is a mechanical force.

Power balance across the transforming factor gives

TF:$k$ $\qquad k.u_m^F.v_m^{Ca.TnC} = u_m^F.v_m^F \qquad (5)$

where $v_m^F(= k.v_m^{Ca.TnC})$ (m.s$^{-1}$), and $k$ (m.mol$^{-1}$) provides the conversion between the mechanical variables and the biochemical variables.

The chemical potentials are given by the Boltzmann relations,

$$u_i^{Ca} = RT \ln(K_{Ca} q_i^{Ca})$$
$$u_m^{TnC} = RT \ln(K_{TnC} q_m^{TnC}) \qquad (6)$$
$$u_m^{Ca.TnC} = RT \ln(K_{Ca.TnC} q_m^{Ca.TnC})$$

and the kinetics for the reaction is

$Re_m^1$ $\quad v_m^{Ca.TnC} = \kappa_1 \left( e^{u_m^{1a}/RT} - e^{u_m^{1b}/RT} \right) = \kappa_1 \left( K_{Ca} K_{Tn} q_i^{Ca} q_m^{Tn} - K_{Ca.TnC} q_m^{Ca.TnC} \cdot e^{-k.u_m^F/RT} \right)$ (7)

where the energy and power balance equations (4, 5) and the Boltzmann relations (6) have been used in the second step in (7). Note that the unbinding rate constant $\kappa_1 K_{Ca.TnC} \cdot e^{-k.u_m^T/RT}$ falls exponentially as the cross-bridge force $u_m^F$ increases.

Substituting (7) into the mass balance equations (3) provides a system of three ordinary differential equations in the three variables $q_i^{Ca}, q_m^{TnC}, q_m^{Ca.TnC}$ and these can be solved subject to specification of initial conditions and the myofilament cross-bridge force $u_m^F$. The functional form of the force dependence of the unbinding rate constant is a direct consequence of formulating the model with bond graphs – i.e., using power balance equations.

Bond graphs have been used to model a wide range of other biological systems. These include metabolic networks [139-141], signaling networks [138], electrophysiological systems [142-144], and mechanochemical systems [145].

The recently developed open-source Python package BondGraphTools allows bond graph models to be easily constructed and simulated [146]. Furthermore, by adopting standards within the CellML framework, bond graph models can be annotated using biological semantics, allowing independently developed models to be automatically merged into larger models [147].

## 5.0 CONCLUSIONS AND OUTLOOK

Many components of the cell processes are expressed as 'regulates', 'involved', 'required' or 'participates' and only give a false impression of complete understanding of a process [148]. With the advent of new technologies delivering more data that is 'big' in multiple ways, we have an opportunity to quantitatively and systematically describe individual cellular processes. These big datasets also present an opportunity to build a unifying cell physiome that mechanistically connects these processes within the natural laws of physics. Further, such a systematic approach will enable predictability across various physiological processes of the cell in its healthy as well as disease states.

We have provided three examples of processes that could use a cell physiome modelling approach, however this is not exhaustive. The cell cytoskeleton is a nexus for all aspects of cell biology that we have not discussed in this review. The cytoskeleton governs overall cell shape [149], nuclear mechanotransduction[150], and acts as the intracellular matrix within which different cell organelles are arranged. Therefore, a cell physiome framework must essentially include a biophysical model of cellular cytoskeletal dynamics and mechanics.

A critical bottleneck in cell physiome modelling is the segmentation of microscopy data for creating 3D computer models for visualization and computational modelling. Additional, compounding factors include readily accessible microscopy data, and computational resources. Nevertheless, we see an exciting opportunity to merge deep learning with cell physiome modelling for parameter estimation of bond graph and other physiome-type models.

Cell physiome modelling is still relatively simplified compared to multi-scale organ modelling [8-12]. Many processes and details of cellular processes are yet to be quantified and encoded mathematically. Bond graphs (and their spatially detailed counterpart port-Hamiltonians) are set to deal with model complexity and conservation of physical laws effectively. However, hybrid stochastic and deterministic models will be key to modelling a wider range of cellular processes. Many sub-cellular events and processes are emergent properties of elementary stochastic events. Endosomal events, at nanoscale, are built on stochastic interactions and processes. Non-thermal active forces and stochastic processes, over time successfully evince a deterministic outcome at macroscale. The intrinsic randomness in the various individual processes justifies incorporating stochastic events into computational models more frequently to understand and predict these yet unknown emergent mechanisms.

Critical to successful adoption of the cell physiome framework will be the access to re-usable, modular computer models and tools. The Physiome Model Repository includes an extensive list of biochemical reaction pathway models. Bond graphs facilitate this approach by providing a unifying framework for coupling models together. We must extend this commitment to include sharing of deep-learning models and 3D geometric models of different cell types in different cellular physiological states. Furthermore, a history of each model needs to be retained so that the relationship to biological data is not obscured over time.

The cell physiome is an ambitious vision, but one that is realistic thanks to the historical efforts in multi-scale modelling of organ systems. Although there are numerous processes that need to be encoded into models, we see this as an exciting opportunity for cross-disciplinary collaboration to systematically put together a quantitative picture of the living cell. A grand challenge of the cell physiome is to mechanistically link nuclear gene transcription to 3D cellular architecture, chemical signals, and mechanical forces.

To predict with mechanistic models is to understand. Through iterative cycles of experimental measurement and modelling, cell physiome modelling will enable us to integrate different datasets and cellular processes to understand and predict cell physiology dynamics.

## Acknowledgements:

The EMBL Australia Partnership Laboratory (EMBL Australia) is supported by the National Collaborative Research Infrastructure Strategy of the Australian Government

|  | Technique | Resolution | Measurement | Reference |
|---|---|---|---|---|
| Big data in number of components | Omics – Metabolomics, transcriptomics, proteomics | Single cells to tissue homogenate | Identity and relative abundance of protein, metabolites, transcripts | [15, 151-153] |
|  | Hyperspectral imaging | Sub-cellular | Multiple organelle dynamics | [18, 154] |
| Big data in time | LLSM/ AOLLSM (sub-seconds to hours) | Sub-cellular | Live cell dynamics, protein distributions | [20, 87] |
|  | Lineage tracing-omics (days) | Single cell | cell state and lineage | [155, 156] |
| Big data in scale | FIB-SEM (>$10^6$ µm$^3$) | nm | Organelle morphology | [157, 158] |
|  | Expansion microscopy (>1 mm$^3$) | nm | Protein distributions, morphology | [21, 22] |
| Big data in characterization | High throughput mechanics | Single cells | forces | [25, 159] |
|  | DNA Nanodevices | Single organelle to cell | pH, membrane potential, forces | [19, 160, 161] |

*Table 1:* A few examples of contemporary techniques of big data in various contexts. This list is not exhaustive.